\documentclass[runningheads]{llncs}
\usepackage{csquotes}
\usepackage[T1]{fontenc}
\usepackage{graphicx}

\usepackage{verbatim}
\usepackage[strings]{underscore}
\usepackage{csquotes}
\usepackage[space]{cite}

\usepackage{textpos}
\usepackage{framed}

\usepackage[english]{babel}
\usepackage{hyperref}
\usepackage{color}

\begin{document}

\title{Streamlining personal data access requests: \\ From obstructive procedures to automated web workflows} 

\titlerunning{Streamlining personal data access requests}
\author{Nicola Leschke \orcidID{0000-0003-0657-602X}
\and Florian Kirsten \orcidID{0000-0002-9202-6640} 
\and \\ Frank Pallas \orcidID{0000-0002-5543-0265} 
\and Elias Grünewald \orcidID{0000-0001-9076-9240}
}
\authorrunning{N. Leschke et al.}
\institute{TU Berlin, Information Systems Engineering, Berlin, Germany \\
\email{\{nl, f.kirsten, fp, eg\}@ise.tu-berlin.de}}
\maketitle              %

\begin{textblock*}{1.5\textwidth}(-3cm, -8cm) %
\begin{center}
\begin{framed}
    \textit{Preprint (2023-05-04) before final copy-editing of an accepted peer-reviewed paper to appear in the\\ Proceedings of the \textbf{23\textsuperscript{th} International Conference on Web Engineering (ICWE 2023)}.}\\ The Version of Record can be found here: \url{https://link.springer.com/conference/icwe}
\end{framed}
\end{center}
\end{textblock*}

\begin{abstract}
Transparency and data portability are two core principles of modern privacy legislations such as the GDPR. From the regulatory perspective, providing individuals (data subjects) with access to their data is a main building block for implementing these.
Different from other privacy principles and respective regulatory provisions, however, this right to data access has so far only seen marginal technical reflection. Processes related to performing \emph{data subject access requests (DSARs)} are thus still to be executed manually, hindering the concept of data access from unfolding its full potential.

To tackle this problem, we present an automated approach to the execution of DSARs, employing modern techniques of web automation. In particular, we propose a generic DSAR workflow model, a corresponding formal language for representing the particular workflows of different service providers (controllers), a publicly accessible and extendable workflow repository, and a browser-based execution engine, altogether providing ``one-click'' DSARs.
To validate our approach and technical concepts, we examine, formalize and make publicly available the DSAR workflows of 15 widely used service providers and implement the execution engine in a publicly available browser extension.
Altogether, we thereby pave the way for automated data subject access requests and lay the groundwork for a broad variety of subsequent technical means helping web users to better understand their privacy-related exposure to different service providers.

\end{abstract}

\section{Introduction} 

Modern privacy regulations such as the GDPR or the CCPA 
comprise a broad variety of obligations to be fulfilled by service providers (\emph{controllers}) processing personal data of individuals (\emph{data subjects}).
Besides concepts broadly recognized in the technical domain -- such as security, data minimization/anonymization, etc. -- this also includes provisions and rights regarding \emph{transparency} and \emph{data portability}. 
Taking the GDPR as a blueprint example for other privacy regulations herein, users have a right to access (RtA), obligating service providers to provide them with \textquote{a copy of the personal data undergoing processing} (Art. 15 (3)) upon request.

This right is considered essential in empowering data subjects to make well-informed and self-sovereign decisions regarding, e.g., which services (not) to use (anymore), which data to share with them, etc. \cite{fialova_data_2014}. 
In addition, the RtA is also an indispensable prerequisite for exercising other privacy rights \cite{murmann_tools_2017}, such as the right to rectification or deletion (Art. 16/17) and also facilitates novel practices of data reuse for individual as well as collective purposes \cite{pallas2020towards,veys_pursuing_2021,pallas_icwe_2022_janus}.

The manifold possible benefits notwithstanding, the RtA is rarely used in practice. Some even say it is \textquote{ignored, inefficient, underused and/or obsolete} \cite[p. 4]{ausloos_shattering_2018}. 
This lack of use can largely be attributed to the fact that request processes as implemented in practice are tedious and error-prone from the perspective of data subjects who, in turn, all too often abstain from pursuing data access or from successfully completing a respective request \cite{bowyer_human_2022}.

To address these shortcomings, we propose to render DSARs more accessible and less obstructive for data subjects by means of web automation and a higher order process automation language explicitly tailored to the specifics of DSARs. 
In particular, we contribute:
\begin{itemize}
    \item a generalized data access process model covering manual, web-based and API-driven DSAR implementations found in practice
    \item a formalized data access request process automation language (DARPAL) facilitating the automation of data access requests for said three cases across a broad variety of service providers
    \item a fully functional prototype
    of a browser-based runtime and a corresponding process repository allowing for \enquote{one-click} data access request, and
    \item an initial corpus of 15 provider-specific, automatically executable DARPAL documents particularly including the most important large-scale service providers and thereby demonstrating the practical viability %
    of our approach.
\end{itemize}

These contributions unfold as follows: In sect. \ref{sec:related}, we introduce relevant background and related work from the legal and process perspective to DSARs as well as on web automation in general. On this basis, we delineate our general approach in sect. \ref{sec:approach}, develop our generalized process model in sect. \ref{sec:process-model}, and subsequently transfer it into our DARPAL language (sect. \ref{sec:darpal}). Prototypical implementations for the runtime and the process repository as well as the initial corpus are presented in sect. \ref{sec:prototype}, followed by a discussion and conclusion (sect. \ref{sec:conclusion}).

\section{Background \& Related Work}\label{sec:related}
Relevant background and related work regards the legal givens for DSARs and respective challenges and shortcomings, real-world implementations of data access processes, and web automation in general. These foundations shall be briefly introduced before delineating our approach.

\subsection{Right to Data Access}\label{sec:related-rta}
At the core of any modern privacy regulation is the right of data subjects to know whether and, if so, what data about them is being processed and by what means.
This includes transparency information that can be obtained before the data is processed (\textit{ex-ante transparency}) \cite{tilt,gruenewald2021tira}, for example through privacy statements, as well as after the data has been processed (\textit{ex-post transparency}) \cite{hildebrandt_behavioural_2009}. 
One important part of ex-post transparency is the RtA, which includes the right to obtain a copy of the personal data undergoing processing.

Under the GDPR, for instance, the RtA is defined in Art. 15. Besides general obligations addressing ex-ante transparency as defined in Art. 15(1), Art. 15(3) declares that data subjects have the right to receive, upon request, a copy of all personal data relating to them and processed by a controller. 
In the following, we will refer to a respective request as \textquote{\emph{data subject access request (DSAR)}}. 
Closely related to the RtA is the right to data portability, as codified in Art. 20 GDPR.
It grants data subjects the right to receive a copy of personal data \textit{they provided} in a \textquote{\emph{structured, commonly used and machine-readable format}}.
Despite this difference of scope, service providers often do not differentiate between data access and data portability \cite[p.~79f]{bufalieri_gdpr_2020} and handle both with the same functionality and interfaces like \textquote{checkout} dashboards or \textquote{archive downloads}. To avoid overly legalese elaborations, we will thus not further distinguish between RtA and RtDP herein but rather refer to them conjointly as RtA and to said interfaces as \textquote{\emph{RtA endpoints}}.

As for the practical implementation of such data access, various challenges have been identified, ranging from controllers not fulfilling their obligations in matters of data completeness \cite{alizadeh_gdpr_2020,bowyer_human_2022,mahieu_collectively_2018} or response time\footnote{The controller has to fulfill the request without delay, but no later than within 30 days that might even be extended up to 60 days, if the data subject is informed about the delay in the first 30 days (Art. 12 (3) GDPR).} \cite{bowyer_human_2022,joris_exercising_2020,urban_study_2019} over a lack of authentication processes, enabling the unauthorized access to personal data \cite{bufalieri_gdpr_2020,cagnazzo_gdpirated_2019,di_martino_personal_2019}, to incomprehensible responses \cite{bowyer_human_2022,mahieu_collectively_2018,veys_pursuing_2021}. 
Last but not least, data subjects often have difficulties locating the RtA endpoint \cite{petelka2022generating}.
Altogether, this severely hinders the execution of the RtA, raising the question of how the process of executing data  requests can be streamlined and made more accessible.

\subsection{Data Access Process}
\label{dsar_wf}
From a data subjects perspective, the execution of a DSAR is composed of two sub-processes: the actual data access process (which we will refer to as DAP), and the privacy enactment, which covers all subsequent activities building upon the data retrieved through the DAP, such as data mapping, visualization, exploration, or (collective) decisionmaking \cite{alizadeh_gdpr_2020,mahieu_collectively_2018,veys_pursuing_2021}. 
However, these enactment activities are rather subject to socio-technical and human-computer-interface research and shall therefore be considered out of scope herein. 
Real-world implementations of DAPs, in turn, can be categorized into one of the following three distinct approaches: 

\textbf{Manual DAPs} require data subjects to identify the responsible data protection officer and make a written (mail, e-mail) or oral (e.g., phone call) request to get a copy of the data, which may either be received via e-mail or as a printed copy \cite{bowyer_human_2022,petelka2022generating,urban_study_2019}.
From the perspective of a data subject, these manual processes cannot be automated, except from using text generation tools that assist in formulating proper request texts.\footnote{Examples of such DAP specific text generators are \href{https://www.datenanfragen.de/}{datenanfragen.de} or \href{https://www.mydatadoneright.eu/}{mydatadoneright.eu}.}

\textbf{Web-based DAPs}, in turn, use a web form as RtA endpoint. In this process, the data subject needs to identify the URL of the form, fill in the form fields and submit the request form.\footnote{Instructions for such processes can for example be found at \href{https://justgetmydata.com/}{justgetmydata.com}.} 
The subsequent communication is usually handled via e-mail, allowing for a download of the retrieved data \cite{urban_study_2019}.
Such a web-based process is somewhat more automated than a manual process, nonetheless, it offers even more automation potential.

Finally, \textbf{API-based DAPs} are characterized by enabling third-party tools to handle authentication, configuration, and request through, e.g., a (standardized) API as proposed in \cite{hansen_generic_2022}, privacy dashboards \cite{schufrin2021}, or personal information management systems (PIMS) \cite{janssen2020personal}.  
Basically, this approach can facilitate tools that fully automate the RtA on the data subject side.\footnote{An API-based DAP is followed, e.g., by the aeon prototype (\href{https://aeon.technology/}{aeon.technology}) that integrates a few big service providers.}
However, such processes are, in practice, rarely implemented \cite{bufalieri_gdpr_2020,urban_study_2019,petelka2022generating}.

\subsection{Web Automation}
As there are many repetitive tasks to handle in the daily interaction with websites, approaches to automate those tasks have been widely discussed \cite{bolin_chickenfoot_2005,cypher1993watch,little_koala_2007}, culminating in the research area of web automation.
Web automation tools can be applied for different purposes and audiences: from allowing developers to automatically test web pages \cite{mickens_mugshot_2010,sharma_web_2014}, over business use-cases involving abstractions of manual work \cite{amershi_liveaction_2013, little_koala_2007}, to improved accessibility, e.g. for visually-impaired users \cite{bigham2009trailblazer,puzis2013predictive}.
The to-be-automated tasks and corresponding actions are wrapped in a so-called macro \cite{cypher1993watch}.
In general, two approaches for generating macros can be distinguished:
declaring a sequence of actions (sometimes referred to as handcrafting \cite{puzis2013predictive}), like, e.g., used in Chickenfoot \cite{bolin_chickenfoot_2005}, and programming by demonstration (PBD) \cite{barman_ringer_2016,cypher1993watch,little_koala_2007}.
The latter approach is more attractive for end-users, like the data subjects in the RtA process, because it requires less \cite{leshed_coscripter_2008} or even no programming knowledge at all \cite{barman_ringer_2016}.
Therefore, we concentrate on PBD, which can be executed in two steps: first, a list of tasks needs to be identified, and then these tasks are automated by using record and replay techniques \cite{chasins_browser_2015}. Those tasks can either be reconstructed by observing user actions \cite{chasins_browser_2015,leshed_coscripter_2008} or application states \cite{cypher1993watch,lau2003programming}.

Above-mentioned traditional PBD mainly addresses single page applications, however, in modern web applications, the to-be-automated tasks might be distributed across web pages.
For that specific case, an intersection with AI-enabled robotic process automation (RPA) \cite{leno2018multi,van_der_aalst_robotic_2018} can be identified, resulting in the research field of web RPA \cite{agostinelli2020automated,dong_webrobot_2022}.

\section{Proposed Approach}\label{sec:approach}
As identified in sect. \ref{sec:related-rta}, current real-world implementations of the RtA broadly lead to data subjects not being able to or abstaining from actually exerting it. 
There is thus a significant need for improving the prevailing status quo of data access. Given their broad use in practice (especially by large-scale service providers) and their already semi-automated nature, we here see particular potential in the automation of web-based DAPs. More specifically, the primary goal behind the work presented herein is to automate the request-parts of DAPs, which are the necessary basis for subsequent data checkouts (for a more detailed analysis of the DAP, see sect. \ref{sec:process-model} below) and to thereby make them more accessible and less obstructive for data subjects. 
In addition, albeit less focused on herein, we also want to pave the way for technically supported DAPs for manual and API-driven approaches across highly variable and provider-specific interfaces.

In a first step, we propose a generalized process model for DAPs that covers manual, web-based, and API-driven DAPs and a broad variety of respective real-world implementations. In particular, this model distinguishes between the sub-processes for submitting a request (data request process, \emph{DRP}) and for checking out the personal data after provision (data checkout process, \emph{DCP}). Of these, we thenceforth primarily focus on the DRP (see above). 

Building upon that process model, we introduce a formal language called DARPAL allowing to represent provider-specific, heterogeneous DRPs -- which we refer to as provider-specific \emph{DRP specifications} herein -- in a unified, formalized, and to-be-executed form. To fulfill the primary goal, such a representation must comprise all automation-relevant parameters, such as the specific RtA endpoint or the to-be-automated workflow itself, as well as the parameters available for customizing the data to be retrieved. In addition, the language shall be defined in such a way that it also covers manual and API-driven approaches to facilitate later extensions in line with the secondary goal.

Based on this language, we aim to make respective provider-specific DRP specifications automatically executable on the data subject side without any intermediary party being involved in the actual DRP execution. In addition, actually executing such DRPs should be as accessible and easy-to-use as possible for data subjects. Employing modern techniques of web automation, we thus strive to provide a browser-based execution engine particularly tailored to the specific givens of DRPs that automates respective requests as far as possible using the parameters provided in a DRP specification. Last but not least, we also aim to make DRP specifications codified in our formal language publicly available and accessible in an extensible process repository and to pre-fill this repository with a base corpus of formalized DRP specifications for broadly used service providers, especially including GAFA\footnote{GAFA represents the four tech-companies Google, Apple, Facebook, and Amazon.}.

Together, these components -- language, browser-based execution engine, process repository, and base corpus -- shall then provide the necessary basis for \enquote{one-click} DRPs. At the same time, successfully automating DRPs for a substantial number of service providers also validates the suitability of our generalized process model and our formal language. All these components shall thus be elaborated on in more detail below.

\section{Modeling Data Access Processes}\label{sec:process-model}
Data subjects wishing to exercise their RtA are following a \textit{data access process (DAP)}, a generalized depiction of which is given in fig. \ref{fig:dap}. 
In general, the DAP can be simplified as a sequential process that can be divided into two sub-processes \cite{petelka2022generating}, which we denote \textit{data request process (DRP)} and \textit{data checkout process (DCP)}, and that can be delineated as follows:

\begin{figure}[t]
    \centering
    \includegraphics[width=\textwidth]{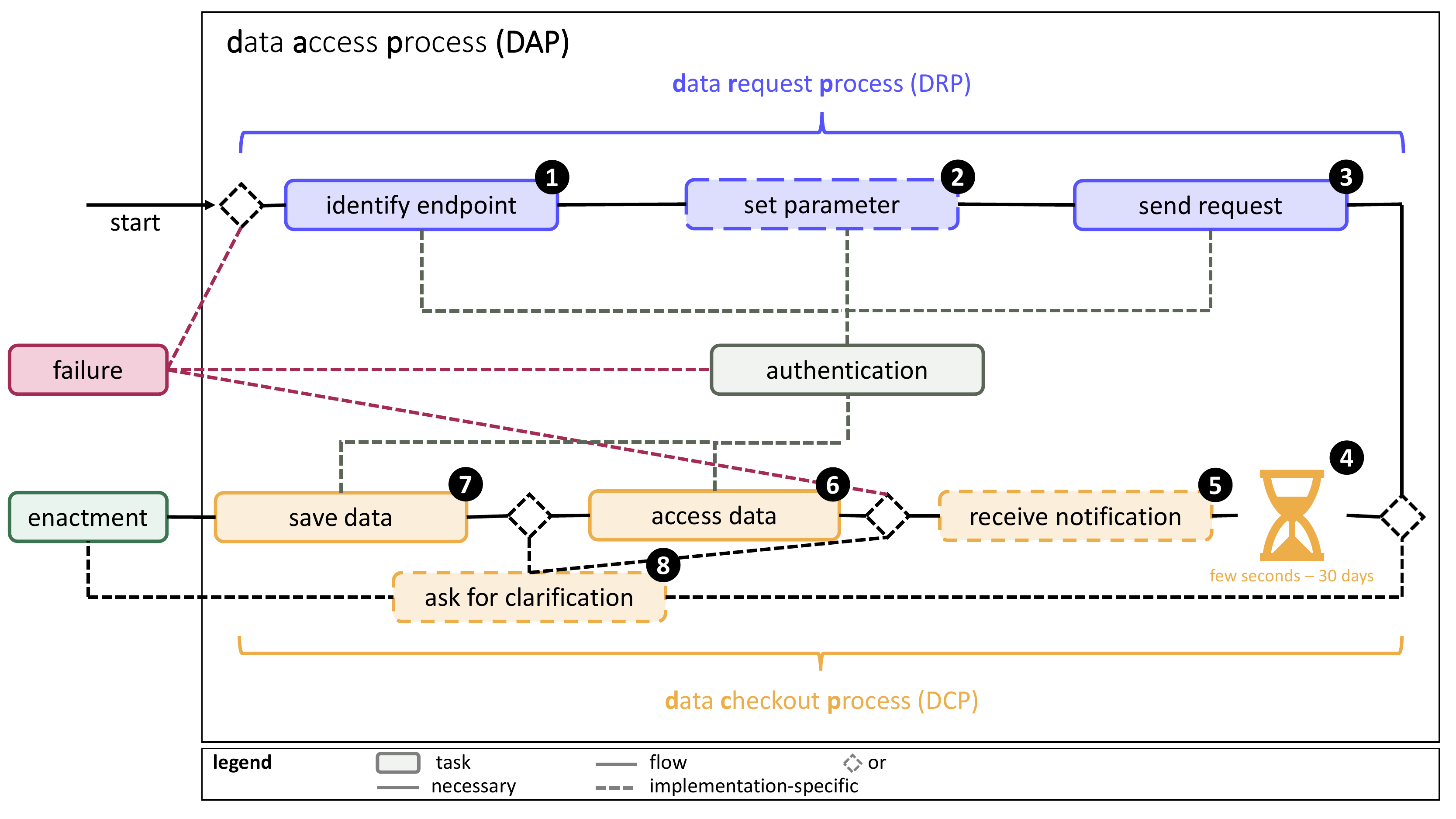}
    \caption{Generic data access process}
    \label{fig:dap}
\end{figure}

\begin{figure}[ht]
    \centering
    \includegraphics[width=0.8\textwidth]{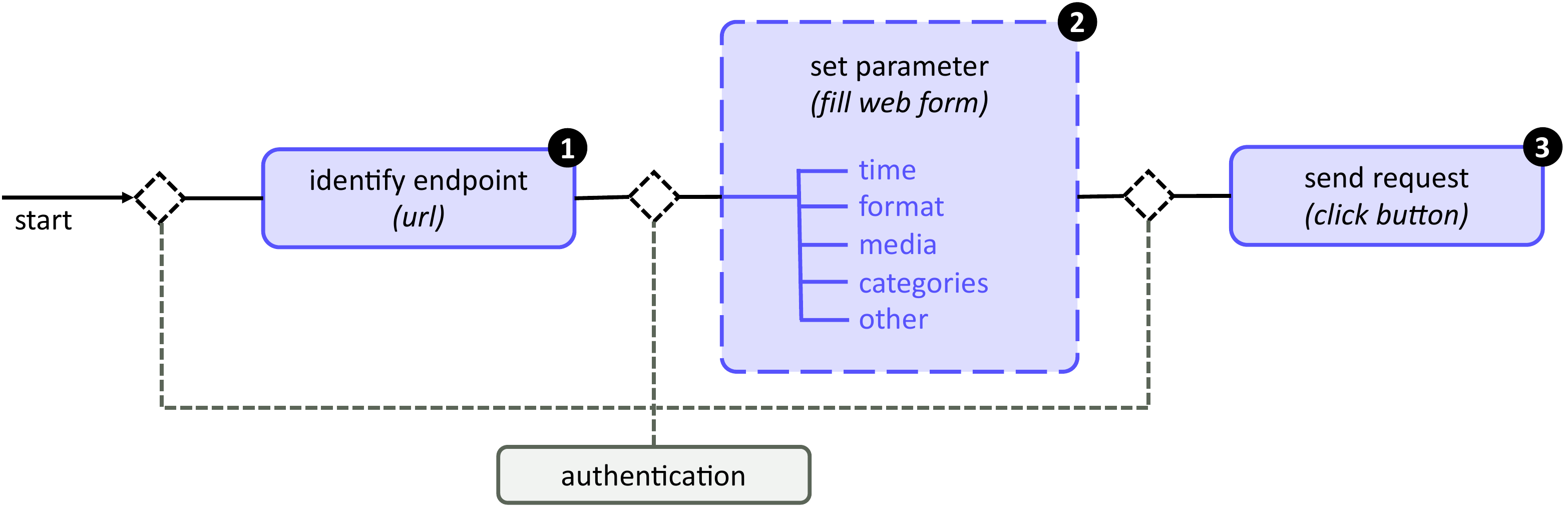}
    \caption{Model of a web-based DRP}
    \label{fig:drp}
\end{figure}

\textbf{Data Request Process (DRP):}
To initiate a DRP, a data subject first has to identify and navigate to the controller-specific RtA endpoint.  
Secondly, some providers require providing further request parameters, e.g., a specific time frame that should be contained in the retrieved data, while others are satisfied by the statement that data access is requested.
Afterwards, the request can be sent.
Of particular interest to reach our primary goal are web-based DRPs, which can be refined as depicted in fig. \ref{fig:drp}.
Within this approach, the RtA endpoint is represented as a web-page or -form, and also the customization is made via a web-form.
In most cases, the web form is only available for authenticated users, so sometimes further authentication processes are dismissed, but in other cases, there are (multiple) authentication steps in place \cite{di_martino_revisiting_2022}.

Even though these steps seem to be quite easy to handle, in reality, the DRP is quite complicated.
For instance, it has been empirically shown that even the very first step is far from straightforward for data subjects, who already struggled in identifying the correct RtA endpoint \cite{alizadeh_gdpr_2020}. This is due to the fact that RtA endpoint addresses are either wrapped in lengthy privacy policies or are de-coupled from the main service experience, e.g., a web form is only accessible via privacy settings hidden in some menu \cite{petelka2022generating}.
 Additionally, the request parameters to be provided can be complex and hard to understand. For example, a non-technical data subject can hardly %
 judge which data format should be used for the download \cite{petelka2022generating}.

\textbf{Data Checkout Process (DCP):}
After the request was successfully made (and, thus, the DRP completed), the DCP starts with a waiting period, the duration of which can range from a few seconds (e.g., Google Takeout) up to the legally specified maximum of 30 days. 
During this period, the service provider processes the request and creates the response.\footnote{Respective provider-side processes are considered out of scope herein.}
Then, some service providers send the compiled data directly to the data subject, while others send a notification that it is available for download \cite{bowyer_human_2022}.
Finally, the data subject can access and save a copy of the data.
Within the DCP, clarification can be required \cite{petelka2022generating} at various stages, arising from both the service provider (see, e.g., \cite{ausloos_shattering_2018}) and the data subject side, especially when above-mentioned challenges with the RtA emerge.

\textbf{Authentication:}
Throughout the whole DAP, multiple authentication steps might be required, which can vary from weak evidence like giving the full name of the data subject \cite{di_martino_revisiting_2022}, logging into a user account or a verified mail account \cite{joris_exercising_2020}, to proving the identity with an ID-card \cite{bowyer_human_2022}. 
Again, data subjects often fail in the authentication steps (or are at least demotivated from proceeding), especially since some authentication processes require multiple clarification messages between the data subject and the service provider, ultimately leading to a DAP not being completed successfully \cite{petelka2022generating}.
However, Joris et al. \cite{joris_exercising_2020} found that only a fraction of service providers included in their study had implemented at least one identification measure. This can lead to an abuse of the RtA, as already mentioned above.

\textbf{End of Process:}
After the DCP has successfully finished, the enactment phase can be started.
In some cases, feedback to the service provider is required, especially if the received data is incomplete. This may lead to a renewed DCP.
Furthermore, the process can also end with an error \cite{petelka2022generating}. Data subjects abort the process at various points because they no longer know what to do \cite{bowyer_human_2022} or because the service provider does not respond (in time) or refuses to process the request \cite{urban_study_2019}.

\section{DARPAL: Process Automation Language} \label{sec:darpal} 

Based on the general process depicted above, we now define a formal automation language specifically tailored to the domain of DRPs. Our \emph{data access request process automation language (DARPAL)} is provided as JSON Schema specification and pays regard to the particularities of DRPs as follows:

First, DARPAL allows specifying the \textbf{requestInterface}, reflecting the RtA endpoint from section \ref{sec:related-rta} above, in a way that covers as many real-world DRP-implementations as possible. In the light of the different DAP categories to be found in practice and their specifics (see section \ref{dsar_wf}), DARPAL thus allows to specify different request interface attributes for manual, web-based, and api-based DRPs. For each of these cases, it must be specified whether it is available and the respective means of authentication (e.g. \enquote{password} or \enquote{id-card}) can be specified. Besides this, the definition of the request interface attributes varies between said three cases:  
In case of a manual process, the interface is an $address$, $email$, or $phone$ number, while for web-based DRPs it is the $startUrl$ of the request form and for APIs it is the $endpointUrl$ and possibly an additional set of $apiParameters$.

For web-based request interfaces -- which we consider the primary target for automating DRPs herein -- a generic second-level field $workflowContainer$ is introduced. It is of type object and has the sub-fields $automationEngine$, $workflow$, $version$, and $verified$.
We decided to provide a generic solution allowing for different automation engines to be used for executing the ultimate workflow. Therefore we need to reference the automation engine.
The workflow itself is also an array of objects specific to the respective automation engine. These objects are not subject to further constraints (except for being a JSON object), because they are of an engine-specific format. The fields $version$ and $verified$, in turn, serve secondary purposes that shall not be discussed in more detail here.

\begin{table}[ht]
\caption{First and second-level building blocks of DARPAL. Mandatory elements are denoted by a *.}\label{tab:interface}
\begin{tabular}{llllll}
\cline{1-2} \cline{5-6}
\textbf{meta}             &              &  &  & \textbf{requestInterface} &                   \\ \cline{1-2} \cline{5-6} 
name*                     &              &  &  & manual                    &                   \\
version*                  &              &  &  &                           & available*        \\
\_hash*                   &              &  &  &                           & address           \\
                          &              &  &  &                           & email             \\ \cline{1-2}
\textbf{requestParameter} &              &  &  &                           & phone             \\ \cline{1-2}
timeRange*                &              &  &  &                           & authentication    \\
                          & allTime*     &  &  & webinterface              &                   \\
                          & customRange* &  &  & \textbf{}                 & available*        \\
mediaQuality              &              &  &  & \textbf{}                 & startUrl          \\
dataFormat*               &              &  &  & \textbf{}                 & authentication    \\
additionalProperties      &              &  &  & \textbf{}                 & workflowContainer \\
                          &              &  &  & api                       &                   \\
                          &              &  &  &                           & available*        \\
                          &              &  &  &                           & endpoint          \\
                          &              &  &  &                           & authentication    \\
                          &              &  &  &                           & apiParameters    
\end{tabular}
\end{table}

Besides the so-structured requestInterface definition, the second concept that needs to be reflected by DARPAL are request parameters, which refine the content of the copy that is to be requested. 
For example, Facebook has a \enquote{download your information} page for logged-in users with a request form that has many options regarding the file format and content of the DSAR, while LinkedIn has a web form where authenticated users have the choice between requesting a full copy or selected general information blocks. Given that these parameters become relevant independently from the automation level of the interface (e.g., the file format can be of relevance for manual DRPs as well), they are specified in a separate \textbf{requestParameter} section. 

Following our model depicted in fig. \ref{fig:drp}, we allow for the fields $timeRange$, describing the period for which the copy is created, $dataFormat$, describing the file type of the copy, $mediaQuality$, describing the preferred quality of media files within the copy, and $categories$ if only specific types of personal data shall be included in the copy. Service-specific other fields can be provided via the $additionalProperties$ field. For example, Linkedin allows to select special $categories$ of personal data that shall be requested. 

For identifying provider-specific DARPAL documents, we additionally need to provide \textbf{meta} information in a third DARPAL section. These meta-information especially include the name of the service provider the document refers to, as well as information about the version and a hash of the document. Together, these three sections make up the DARPAL document structure depicted in table \ref{tab:interface}. Additionally, there are some fields that are required for document identification, however, they have no further semantic meaning and are therefore not elaborated here. The respective JSON-Schema definition also providing more details on field formats / data types etc. is available online.\footnote{\href{https://github.com/DaSKITA/darpal}{github.com/DaSKITA/darpal}}

\section{Prototypical Implementation} \label{sec:prototype}
\begin{figure}[t]
    \centering
    \includegraphics[width=0.8\textwidth]{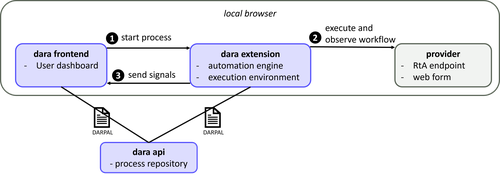}
    \caption{Architecture of the DARA system}
    \label{fig:dara}
\end{figure}

To make DRPs as accessible as possible, we implemented the \emph{\enquote{data access request assistant (DARA)}} system, consisting of a process repository component pre-filled with 15 real-world DARPAL specifications\footnote{\href{https://github.com/DaSKITA/darpal-documents}{github.com/DaSKITA/darpal-documents}}, an automation engine in the form of a publicly available browser extension and a user interface, altogether providing fully functional \enquote{one-click} DRPs.
A general overview can be found in fig. \ref{fig:dara}. Each of these components shall be elaborated on in more detail below.

\textbf{Process Repository:}
The storage and processing of the available DRP specifications is done in a separate process repository, which we refer to as DARA API\footnote{\href{https://github.com/DaSKITA/dara-api}{github.com/DaSKITA/dara-api}}.
In addition to providing automation engines with the necessary DRP specifications, the process repository is also conceptualized (but not yet implemented) to handle quality assurance and lifecycle management of the stored DARPAL documents\footnote{For instance, statistics on successful and failed local workflow executions shall in future versions be reported back to the repository. With this data, likely outdated and dysfunctional workflows could be marked respectively.}.
It was realized as Python-based REST-API via FastAPI. 
For each covered service provider, it offers a separate endpoint with GET, POST and DELETE routes.

Currently, the DARA API includes a base corpus of 15 DARPAL specifications, describing the DRPs of broadly used service providers. These are retrieved and employed by both, the DARA extension and the DARA frontend. The workflows within the specifications are enriched with signals reporting the status of the DAP to the frontend.
For example, we send a \enquote{started execution} event to the execution environment for proper handling as soon as the provider's DRP page was loaded successfully and no redirection to a login portal occurred. 
Statistics collection and crowdsourcing functionalities were not yet implemented in the current prototype, however, they are planned for future versions.

\textbf{Browser Extension:}
For our prototype, we chose an approach based on a browser extension because of the accessible and cross-platform installation procedure through the browsers' extension stores. To execute the DRP in the user's browser, we built the DARA extension\footnote{\href{https://github.com/DaSKITA/dara-extension}{github.com/DaSKITA/dara-extension}} based on the Automa \footnote{\href{https://www.automa.site/}{www.automa.site}} 
project. The Automa automation engine uses a simple block-based user interface to build browser workflows. We configured Automa to include a content-script in our frontend, which listens for user commands like the execution of a particular DRP or, respectively, the workflow it includes.

When an automated DRP is triggered, either directly via the extension or via the frontend, the DARA extension executes it in a new browser tab, which is automatically opened in the background. 
After successful execution, this tab is closed and a \enquote{success} signal is sent to the frontend. However, if the execution fails or takes longer than expected, an \enquote{interaction required} signal is sent, and the user can switch to the background tab.
To additionally enable non-technical users to add or update (individual, local) DRP specifications, we build upon the workflow recording functionality of Automa. We modified the recording functionality so that per default, XPath is used to select HTML elements, as in our experience XPath provided a more stable selector than using CSS classes.

\textbf{Frontend:} 
Via a dashboard\footnote{\href{https://github.com/DaSKITA/dara-frontend}{github.com/DaSKITA/dara-frontend}}, the user is first asked to install the DARA extension. Afterwards, the user assesses the DRP specifications available via the DARA API and selects the service provider for which a DRP shall be executed.
While executing a DRP in a non-active tab through the web extension, the user is informed about the execution state and in case manual interaction is required, the option to switch to the browser-tab in question is displayed. Currently, the frontend is available in German, with further languages to be added soon.

\textbf{Preliminary User Study:}
In a preliminary pre-study, we asked 14 users (not familiar with our implementation before) to request their data from two of the fifteen service providers we provide DARPAL specifications for.
For the first chosen service, the users had to follow the DRP without the use of our prototype or any other instructions. It took them on average 4.3 minutes to finish the DRP. 29\% of the users were not able to successfully send a request at all. Overall, only half of the user group considered executing their RtA again. 
The second service was requested using our prototype. The users chose to send a request to Amazon, Apple, Google, Instagram, LinkedIn, and Vinted. In all cases, it took them less than two minutes to complete the request (with an average of one minute), and 86\% of the users stated they are willing to repeat the process on a regular basis, with the same and other service providers.

Even though being of highly preliminary nature and hardly generalizable so far, these results are definitely encouraging and at least support the basic viability of our approach. 
Following up on the more technical groundwork presented herein, a more in-depth user study is planned in the near future.

\section{Discussion \& Conclusion}\label{sec:conclusion}
With our contributions as a whole, we provide significant improvements for data subjects exercising their RtA. Formerly tedious web-based data requests are condensed into automated \enquote{one-click} DRPs.   
Still, the approach pursued herein carries inherent needs for creating a multitude of provider-specific DARPAL documents separately. Our prototype partially tackles this problem with a dedicated recording function. 
Still, service providers immensely obfuscate their page sources, perform random changes to the document object model, or follow other strategies preventing effective web automation. 
Future versions of our prototype aim to counteract this problem with crowdsourcing features. Nonetheless, such options also include abuse potentials (spamming, security and trust issues etc.). Thwarting these remains a nontrivial challenge. Perspectively, DRP automation could substantially profit from documented APIs, which unfortunately exist far too rarely so far. We, thus, emphasize the regulatory need for documented APIs \cite{gill2022data}, optimally with queryable metadata according to Art.~15(1)~GDPR.

Further developments in DARPAL could include a generic process representation, which could then be consumed by different automation engines and translated into their specific representation before execution. For the moment, however, we used Automa as the mere engine and workflow format and did not run into relevant limitations. 
For greater interoperability, however, others should be supported as well.

We encourage data subjects to request their data from as many services as possible. 
However, we solely addressed the DRP herein, while the DCP and enactment processes are not yet included. Still, they are very important to make access requests useful. In particular, the received data packages need to be prepared for actual interpretation, summary, risk identification, or visualization. Only with such data post-processing, data subjects can actually make sovereign evaluations and decisions regarding their usage and the processing activities of a service offering \cite{veys_pursuing_2021}. Respective functionalities are natural candidates for extending our solution further in the future. This would also be in line with requests from two of our study participants explicitly asking for data retrieval and analysis support to also be provided. 
Still, our approach shall serve as stable ground for respective upcoming work dealing with the actual data presentation tasks needed for a meaningful enactment stage.

Another promising next step for fostering the RtA across platforms is to extend the scope from webpage-based services to, e.g., native mobile apps (for instance through deep-linking). Similarly, our system could also be integrated with Personal Information Management Systems (PIMS) or Personal Data Stores (PDS) \cite{janssen2020personal}. These aim to manage granular data access based on, among others, purpose-specific consent provisions for selected categories of personal data. Our contributions can be used to import such data, possibly also to perform data transfers in the vein of the RtDP. 

Altogether, the contributions presented herein thus provide a solid ground for much more structured, automated, and sovereign DRPs, illustrating that and how web automation can fill a gap left open by existing regulations. This can not only have an instant effect on the actual exercise of the RtA, but also pave the way for a broad variety of future research endeavors in the context of the RtA, the RtDP, and technically mediated ex-post transparency in general.

\subsubsection{Acknowledgements.}
\begin{small}
We thank our students Majed Idilbi, Christopher Liebig, Ann-Sophie Messerschmid, Moriel Pevzner, Dominic Strempel, and Kjell Lillie-Stolze, who contributed to the initial proof-of-concept within the scope of a study project \cite{datensouv2021}. Special thanks go to Johanna Washington, who kindly supported us to conduct the user experiment.

The work behind this paper was partially conducted within the project DaSKITA, supported under grant no. 28V2307A19 by funds of the Federal Ministry for the Environment, Nature Conservation, Nuclear Safety and Consumer Protection (BMUV) based on a decision of the Parliament of the Federal Republic of Germany via the Federal Office for Agriculture and Food (BLE) under the innovation support program.
\end{small}

\bibliographystyle{splncs04}
\bibliography{references}

\end{document}